\documentclass{article}
\usepackage{graphicx}

\PassOptionsToPackage{numbers, compress, sort}{natbib}

\usepackage[final]{neurips_2023_ml4ps}

\usepackage[utf8]{inputenc} 
\usepackage[T1]{fontenc}    
\usepackage{hyperref}       
\usepackage{url}            
\usepackage{booktabs}       
\usepackage{amsfonts}       
\usepackage{nicefrac}       
\usepackage{microtype}      
\usepackage{xcolor}         
\usepackage{subcaption}
\usepackage{multirow}

\title{Efficient and Robust Jet Tagging at the LHC with Knowledge Distillation}

\author{%
  Ryan Liu\\
  University of California, Berkeley\\
  Berkeley, CA 94720\\
  \And
  Abhijith Gandrakota\\
  Fermi National Accelerator Laboratory\\
  Batavia, IL 60510 \\
  \And
  Jennifer Ngadiuba\\
  Fermi National Accelerator Laboratory\\
  Batavia, IL 60510 \\
  \And
  Maria Spiropulu\\
  California Institute of Technology\\
  Pasadena, CA 91125\\
  \And
  Jean-Roch Vlimant\\
  California Institute of Technology\\
  Pasadena, CA 91125\\
}

\begin{document}

\maketitle

\begin{abstract}
    The challenging environment of real-time data processing systems at the Large Hadron Collider (LHC) strictly limits the computational complexity of algorithms that can be deployed. For deep learning models, this implies that only models with low computational complexity that have weak inductive bias are feasible. To address this issue, we utilize knowledge distillation to leverage both the performance of large models and the reduced computational complexity of small ones. In this paper, we present an implementation of knowledge distillation, demonstrating an overall boost in the student models' performance for the task of classifying jets at the LHC. Furthermore, by using a teacher model with a strong inductive bias of Lorentz symmetry, we show that we can induce the same inductive bias in the student model which leads to better robustness against arbitrary Lorentz boost. 
\end{abstract}

\section{Introduction}
In the past decades, deep learning has transformed the data analysis workflow at high-energy physics experiments, achieving state-of-the-art performance on many challenging tasks such as jet tagging, charged particle tracking, particle flow algorithm, and neutrino event reconstruction \cite{qu2022particle, gnn4itk, Pata_2023, shmakov2023interpretable}. Such breakthroughs mostly follow from novel architectures and large-scale datasets. In particular, architectural innovation involves both general-purposed models adopted from deep-learning research (e.g. convolutional neural networks, transformers, graph neural networks \cite{cms2020identification, qu2022particle, shlomi2020graph}) and custom models that incorporate strong inductive biases \cite{gong2022efficient, bogatskiy2020lorentz}. However, these paradigms are not applicable to each stage of the experimental data processing workflow, especially to online applications such as the hardware trigger systems~\cite{CERN-LHCC-2017-020, CERN-LHCC-2020-004} of particle detectors at the LHC that are characterized by strict latency, power, and resource constraints. For such applications, a generic method that can improve deep learning models' performance and robustness without scaling the computation in the production environment becomes crucial.

Since knowledge distillation (KD) was proposed by Hinton in 2015 \cite{hinton2015distilling}, it has been widely used in natural language processing applications \cite{jiao2019tinybert, sanh2019distilbert}. KD is a technique to transfer the learned knowledge from a large and cumbersome ``teacher'' model to a more efficient ``student'' model. This is done by replacing the ground truth (hard targets) with teacher-predicted probabilities (soft targets). Remarkably, KD can significantly boost the performance of the student model without any modification to the architecture itself. In this paper, we describe an implementation of KD for jet tagging at the LHC and demonstrate that powerful deep-learning models can be deployed to early-stage event selections. 

\section{Related Work}
\subsection{Group Invariant Neural Networks}
Inductive bias is a crucial part of designing a neural network. In the context of jet physics, we have two important symmetries that shall be taken into consideration: permutation invariance and Lorentz group invariance. Whereas the constituents of a jet come in no particular order, the prediction should not change upon the exchange of two particles in the input point cloud. At the same time, since the source of the jet can have high momentum and cause the jet to be boosted by a Lorentz transformation, the prediction should not change by an arbitrary Lorentz transformation as well.

To incorporate permutation symmetry, it was proven that all permutation invariant functions can be written as follows \cite{zaheer2018deep}:
\begin{equation}
    f(X) = \rho(\sum_{x\in X}\phi(x))
\end{equation}
We can parameterize $\rho$ and $\phi$ using multi-layer perceptrons (MLPs) to make the network learnable. The resulting architecture is called a \textit{deep set} model. To improve the expressiveness of the group invariant neural network, we can also consider attention-based models such as Transformers or Graph Attention Networks \cite{vaswani2017attention, velivckovic2017graph}.

As for Lorentz symmetry, it is typically harder to incorporate its inductive bias. Existing solutions include restricting the message-passing mechanism of graph neural networks to use only Lorentz scalars such as $\langle x, y\rangle$ or $\Vert x - y\Vert^2$ \cite{gong2022efficient} and manipulating representations of $SO^+(1,3)$ by tensor-product and Clebsch–Gordan decomposition \cite{bogatskiy2020lorentz}.

\subsection{Knowledge Distillation}
One of the common problems of the models described in the previous section is that most of them do not scale well, especially for models that require pair-wise computation such as Transformers and Graph Neural Networks. Therefore, we propose to use KD to leverage the power of these models while not adding to the inference-time cost. KD transfers learned knowledge by replacing the hard target normally used to train the small and fast student model with soft targets from the large and cumbersome teacher model. To be more precise, the KD loss is defined as follows:
\begin{equation}
    L_{KD}(q; p, y)=(1-\lambda) \mathcal H(y, q) + \lambda D_{KL}(\tilde p\Vert \tilde q)
\end{equation} 
where $q$ is the student's output probabilities, $p$ is the teacher's output probabilities, and $y$ is the ground truth label. Probabilities with $\tilde \cdot$ are the distributions softened by temperature $T$, i.e., $p(x) = \frac{e^{s(x)}}{\sum_{x'}e^{s(x')}}$ means $\tilde p(x) =  \frac{e^{s(x)}/T}{\sum_{x'}e^{s(x')/T}}$. The first term $H(q, q)$ is the cross entropy loss with the ground truth as the target, and the second term $D_{KL}(\tilde p\Vert \tilde q)$ is the Kullback–Leibler divergence with the softened teacher as the target. It was pointed out that the KD loss has two major effects, namely, it can inject the class relationship prior to the student and rescale the gradients of samples according to the teacher's relative confidence \cite{tang2021understanding}. Moreover, recent studies showed that KD is capable of transferring inductive bias \cite{abnar2020transferring}. That is, if the teacher has a strong inductive bias, the student can learn from the KD loss to generalize the same way as the teacher does. 

\section{Experiments}

\subsection{Top-quark jets Tagging at the LHC}
In high energy physics, a \textit{jet} refers to a collimated shower of particles that result from the decay and hadronization of quarks $q$ and gluons $g$. The goal of \textit{jet tagging} is to identify which mother particle originated the jet such to distinguish interesting and rare signal events from highly frequent background ones. In this paper, we study the effect of KD on a common benchmark dataset for jet tagging, the top tagging dataset \cite{kasieczka_gregor_2019_2603256}. The dataset contains signal top-quark jets and background light quark and gluon jets simulated with Pythia8 \cite{SJOSTRAND2015159} and passed through a simulation of a typical LHC particle detector obtained with Delphes \cite{de2014delphes}. There are 1.2M training events, 400K validation events, and 400K testing events. Each jet is described by its constituents' four momenta, together with a flag indicating if it is a signal event. 

\subsection{Model Architecture}
In this paper, we demonstrate the efficacy of KD on two student models, the deep set model \cite{zaheer2018deep} and an MLP model. The inputs are the constituents' momenta in cylindrical coordinates $(p_T, \eta, \phi, m)$. Here we appended the particle's true mass $m$ to the input such that the model can distinguish between different particle species. Note that the set of inputs we used here are also available in a typical real-time data processing system thanks to the particle flow algorithms at the Phase 2 CMS detector \cite{CERN-LHCC-2017-013}. For the deep set model, we used a 3-layer MLP with a hidden size of $128$ to parameterize both $\rho$ and $\phi$, together with batch normalization \cite{ioffe2015batch} applied after leaky relu activations \cite{xu2015empirical}. The aggregation follows the design of the Energy Flow Network such that per-particle embeddings are aggregated according to their $p_T$ to enforce IR-safety~\cite{komiske2019energy}. For the MLP model, we sort particles according to their $p_T$. Any jet with more than 128 particles is trimmed to 128 particles while jets with less than 128 particles are zero-padded. This gives $4\times 128=512$ input features in total. We then feed them into an MLP of 3 layers with 512 hidden features each, again equipped with leaky relu and batch normalization. For more information, the code of this work is available in this \href{https://github.com/ryanliu30/KD4Jets.git}{Github repository}

\subsection{Experiment Setup}

To understand if KD can improve student performance and transfer inductive biases, we picked the LorentzNet \cite{gong2022efficient} as the teacher model. LorentzNet has a very strong inductive bias of Lorentz-transformations invariance and is one of the best-performing models on the top-tagging dataset. We designed two experiments; (1) we evaluated the efficacy of KD for both the deep set and the MLP models and compared their performance to the ones trained from scratch. We experimented with temperatures of $T\in\{1, 3, 5\}$. (2) To show how KD can transfer inductive bias, we trained the deep set model on an augmented dataset that was boosted by $\beta$ uniformly sampled from $[0, \beta_{max}]$ along the $x$-axis and evaluated the models on the boosted jets test set. We also chose $\lambda = 1$ for this experiment to prevent the model from learning from the ground truth labels. We hypothesized that the ability for KD to transfer inductive biases would be further improved when the student is exposed to some corrupted samples that correspond to the teacher's inductive bias. For each model, we trained for 100 epochs with \texttt{AdamW} optimizer and a \texttt{StepLR} learning rate scheduler. 
\subsection{Results}
We report the accuracy, the area under curve (AUC), and the background rejection ($\mathrm{R}_{X\%}$) of these models on the test set. The background rejection is defined as $1/\mathrm{FPR}$ when the $\mathrm{TPR}$ is fixed to $X\%$. Furthermore, with an eye to deployment to real-time system, we measured the number of floating point operations (FLOPs) with the package \texttt{fvcore} \cite{fvcore}, which computes a model's FLOPs with torch jit tracing.
For the first experiment, the results are reported in Table \ref{test}. Both the deep set and MLP models showed performance gains from knowledge distillation. Remarkably, the overall accuracy improvement for the MLP model is 1.5\%, together with about a factor of two improvement in background rejection. At the same time, the deep set model showed about a 25\% improvement in background rejection. The results demonstrate the effectiveness of knowledge distillation in jet tagging.
\begin{table}
  \caption{Comparison between models trained from scratch and knowledge distillation.}
  \label{test}
  \centering
  \small
\makebox[\textwidth]{\begin{tabular}{lllllll}
    \toprule
         & \#params & FLOPs & Accuracy & AUC & $\mathrm{Rej}_{30\%}$ & $\mathrm{Rej}_{50\%}$ \\
    \midrule
    DeepSet from scratch & \multirow{4}{*}{68.2K}& \multirow{4}{*}{1.67M} & 0.930 & 0.9808 & 747 & 219 \\
    DeepSet KD $T=1$ &&& \textbf{0.932} & 0.9818 & 926 & 241\\
    DeepSet KD $T=3$ &&& \textbf{0.932} & \textbf{0.9819} & \textbf{970} & \textbf{255}\\
    DeepSet KD $T=5$ &&& \textbf{0.932} & \textbf{0.9819} & \textbf{970} & 248\\
    \midrule
    MLP from scratch & \multirow{4}{*}{527K} & \multirow{4}{*}{529K} & 0.904 & 0.9663 & 256 & 82\\
    MLP KD $T=1$ &&& 0.914 & 0.9726 & 375 & 119\\
    MLP KD $T=3$ &&& 0.918 & \textbf{0.9751} & 483 & 144\\
    MLP KD $T=5$ &&&\textbf{0.919} & 0.9750 & \textbf{503} & \textbf{146}\\
    \midrule
    LorentzNet (teacher) &224K& 339M& 0.942 & 0.9868 & 2195& 498  \\
    \bottomrule
  \end{tabular}}
\end{table}
\begin{figure}
    \centering
    \begin{subfigure}[b]{0.4\textwidth}
        \centering
        \includegraphics[width=\columnwidth]{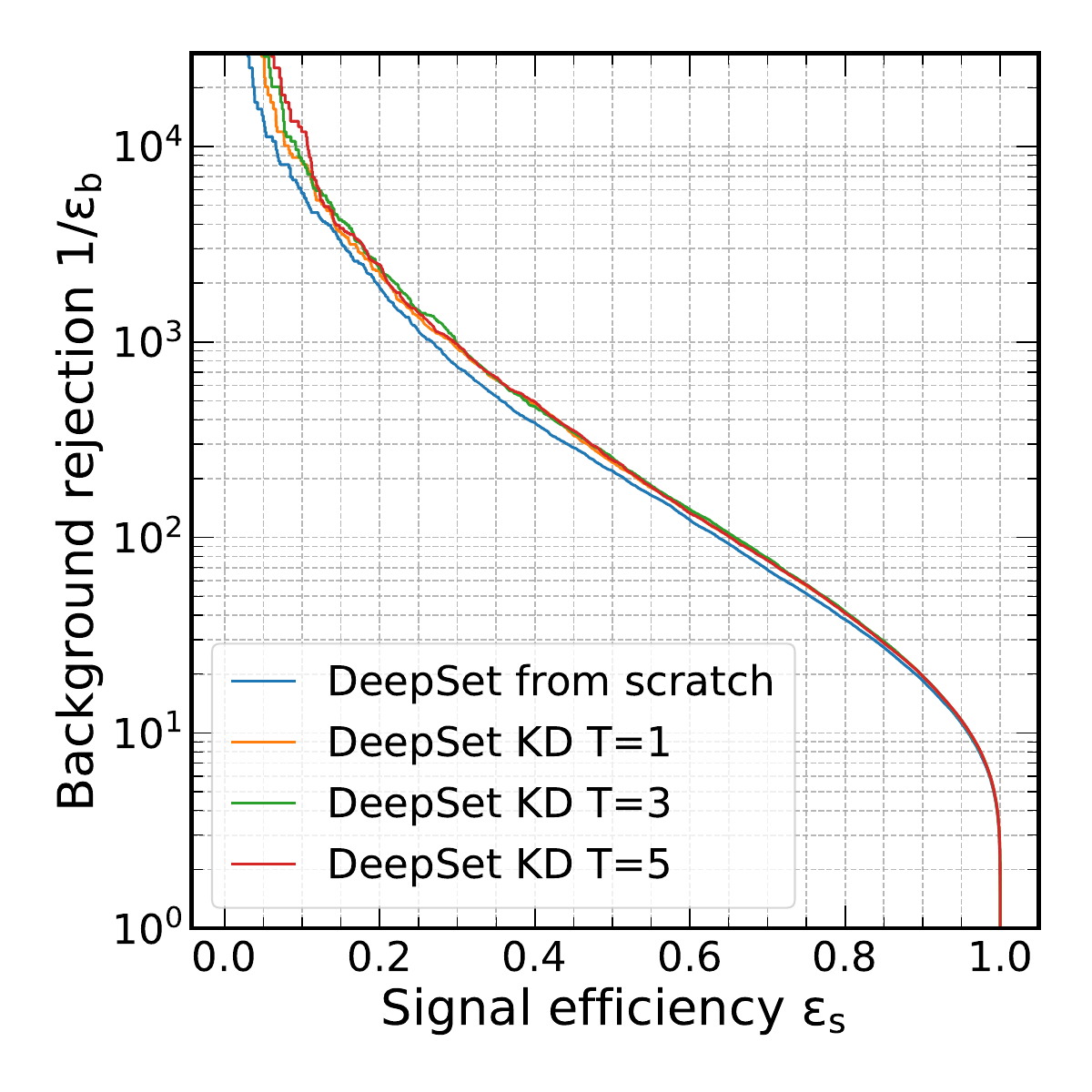}
        \caption{ROC curve of Deep sets}
        \label{fig:roc_deepsets}
    \end{subfigure}
    \begin{subfigure}[b]{0.4\textwidth}
        \centering
        \includegraphics[width=\columnwidth]{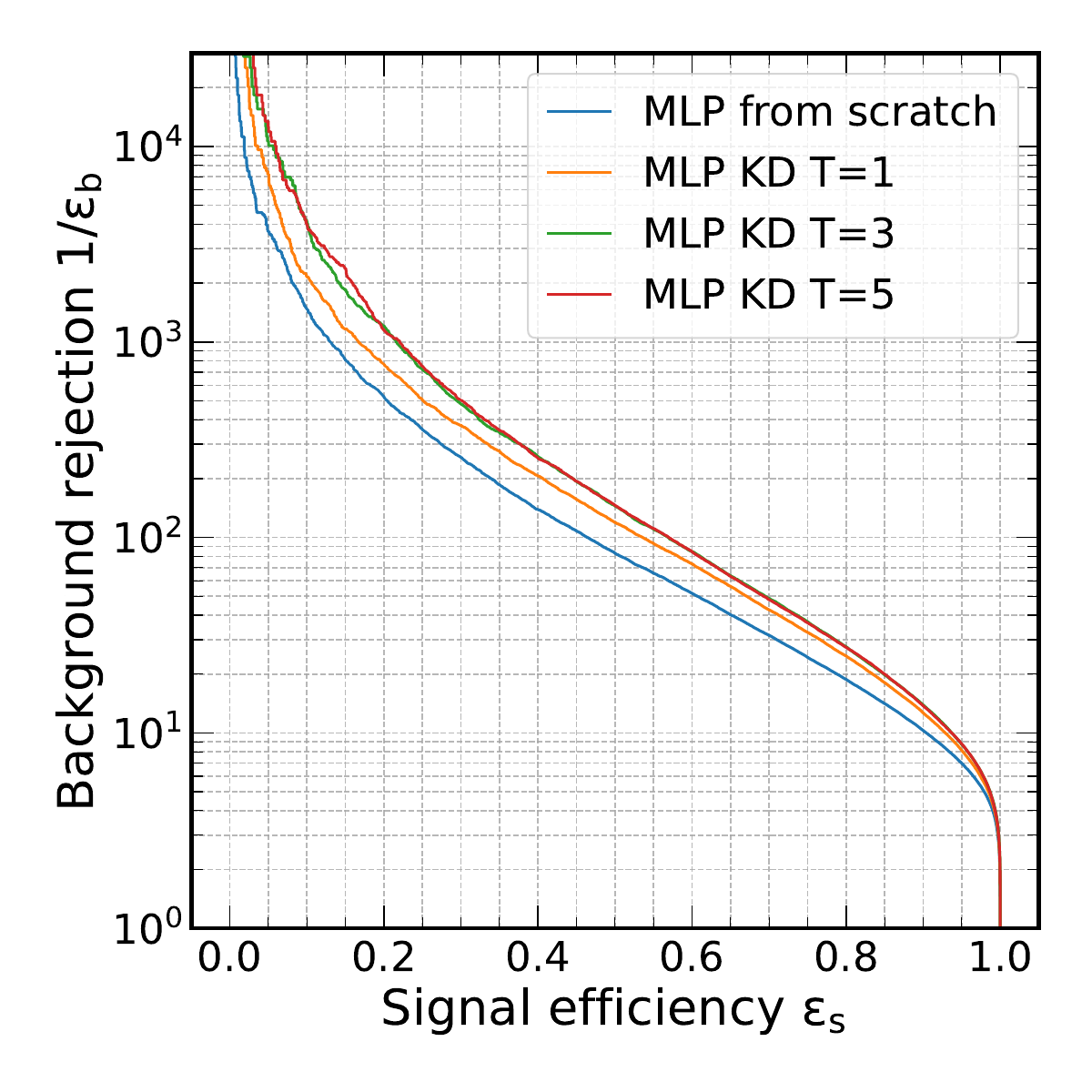}
        \caption{ROC curve of MLPs}
        \label{fig:roc_mlp}
    \end{subfigure}
    \caption{ROC curves of the two deep set and MLP models on the test dataset. The knowledge distillation models have their curves above the models trained from scratch.}
    \label{fig:roc}
\end{figure}

\begin{figure}
    \centering
    \begin{subfigure}[b]{0.45\textwidth}
        \includegraphics[width=\columnwidth]{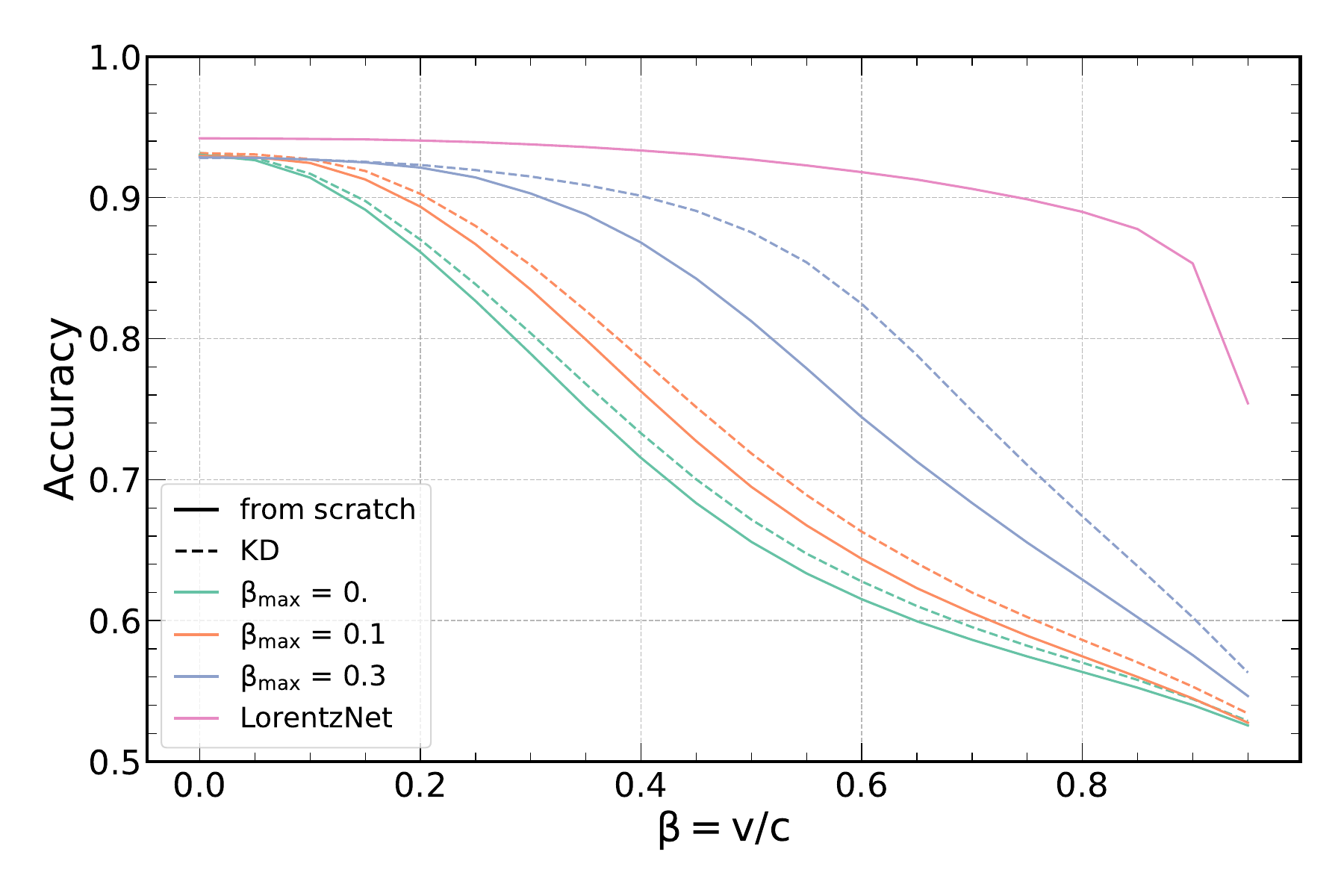}
    \caption[Caption for LOF]{Lorentz invariance test for the deep set model\protect\footnotemark.}
    \label{fig:boost}
    \end{subfigure}
    \begin{subfigure}[b]{0.45\textwidth}
        \centering
        \includegraphics[width=\columnwidth]{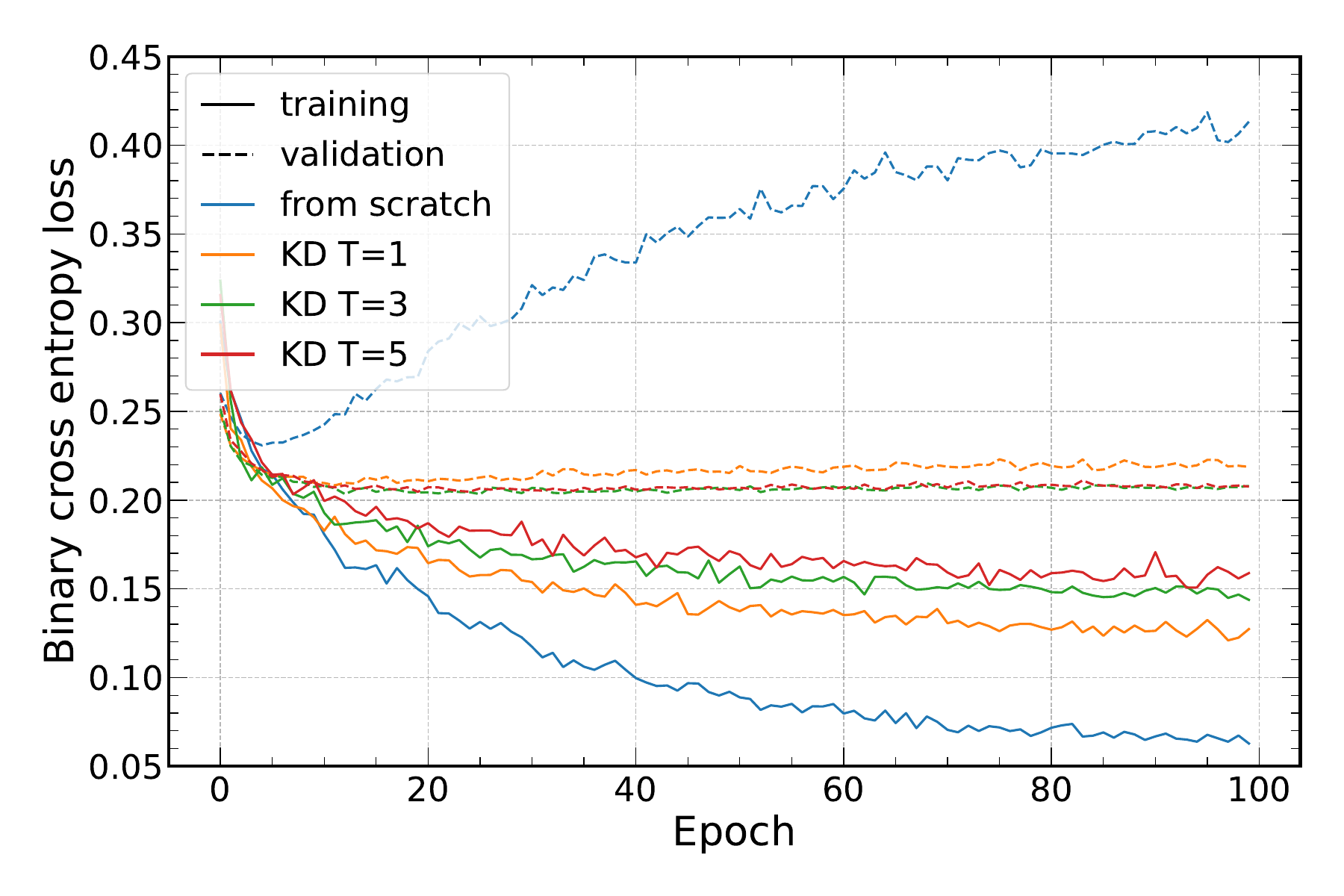}
        \caption{Training hard target loss}
        \label{fig:train}
    \end{subfigure}
    \caption{(a) we can see that the robustness with respect to Lorentz boosts improved with KD in all three augmentation configurations. (b) In the first experiment, the validation loss of MLP models with KD converges instead of increasing as in the case without KD.}
    \label{fig:overfit}
\end{figure}

For the second experiment, in all three configurations, $\beta_{max}\in \lbrace 0, 0.1, 0.3\rbrace$, we can see in Figure \ref{fig:boost} that the robustness of the deep set models significantly improved, presumably due to the inductive biases transferred. 

Finally, we observed that knowledge distillation can prevent models from overfitting the data (see Figure \ref{fig:train}). This is especially useful when the sample size is small. We hypothesize that this is due to a more complicated learning objective that forces the model to generalize instead of memorize.

\footnotetext{The curve here for LorentzNet is different from Figure 3 in \cite{gong2022efficient}. This is because LorentzNet included two auxiliary inputs that represent the beam line. In their paper, the authors also boosted the beams. However, we believe that boosting the beams encodes the information about the boosting so we decided not to follow the convention.}

\section{Conclusion}
In this paper, we showed that knowledge distillation can improve jet tagging performance and robustness. Without adding any additional operation, we realized a $25\%$ improvement in background rejection. Furthermore, we demonstrated that the teacher's inductive bias can be transferred and help the student generalize better. We hope that this work can serve as a base for future deployment of deep learning models to real-time event selection systems at the LHC, bringing the power of deep learning to the frontier of experimental high-energy physics. 

\section{Broader Impact Statement}
We expect that this work will stimulate the research and further discussions on:
\begin{enumerate}
    \item \textbf{Knowledge distillation between different architectures}. Knowledge distillation enables us to leverage the recent breakthroughs in large models to enhance the small models we have. As shown in this work, knowledge distillation can work for models of different architectures.
    \item \textbf{Knowledge distillation as a method to improve robustness}. In the Lorentz invariance test, we have seen that knowledge distillation is capable of improving student's robustness. This may suggest that we can train a cumbersome teacher model on an augmented dataset and transfer the robustness to the student by knowledge distillation without having the student trained on the augmented dataset, avoiding the accuracy and robustness trade-off.  
\end{enumerate}
\section{Limitations}
There exist a few limitations of this work. Firstly, we have only shown the efficacy of knowledge distillation for a classification problem. Whether it would work for other types of tasks in high-energy physics remains unclear. Secondly, training the teacher model may be very costly. In this work, we used a pre-trained model as the teacher. However, if one would like to train a teacher model from scratch the cost of training may be significant. 
\section{Acknowledgement}
This work is listed in Fermilab Technical Publications as FERMILAB-PUB-23-748-CMS.
AG and JN are supported by Fermi Research Alliance, LLC under Contract No. DE-AC02-07CH11359 with the Department of Energy (DOE), Office of Science, Office of High Energy Physics. 
JN and RL are also supported by the U.S. Department of Energy (DOE), Office of Science, Office of High Energy Physics ``Designing efficient edge AI with physics phenomena'' Project
(DE-FOA-0002705). JN is also supported by the DOE Office of Science, Office of Advanced Scientific Computing Research under the ``Real-time Data Reduction Codesign at the Extreme Edge for Science'' Project (DE-FOA-0002501). This work was supported in part by the AI2050 program at Schmidt Futures (Grant G-23-64934).
\bibliographystyle{unsrt}
\bibliography{ref.bib}

\end{document}